\documentclass[12pt,twoside,a4paper]{article}

\usepackage{amssymb,amsmath}
\usepackage{amsfonts}
\usepackage{indentfirst}
\usepackage{graphicx}
\newcommand{\beq}{\begin{equation}}
\newcommand{\eeq}{\end{equation}}

\newtheorem{theorem}{Theorem}

\newtheorem{lema}{Lemma}
\newcommand{\barr}{\begin{array}}
\newcommand{\earr}{\end{array}}

\begin{document}
{\Large\bf Exact treatment of linear difference \\ equations with
noncommutative coefficients} \vspace{0.5cm}
\begin{quote} {\small  \begin{center}M. A.
Jivulescu, A. Messina, A. Napoli  \end{center}
 \textit{MIUR, CNISM and Dipartimento di Scienze Fisiche ed
Astronomiche,\\ Universit\`{a} di Palermo, via Archirafi 36, 90123
Palermo, Italy} \begin{center} F. Petruccione \end{center}
\textit{School of Physics, University of KwaZulu-Natal, Private
Bag X54001, Durban 4000, South Africa }}\end{quote}

\begin{center}
\textmd{SUMMARY}\end{center} The exact solution of a Cauchy
problem related to a linear second-order  difference equation with
constant noncommutative coefficients is reported.

 \begin{center}\textmd{1. INTRODUCTION}\end{center}
\par Difference equations are ubiquitous in applied sciences. The
mathematical theory underlying the treatment of these equations is
well consolidated   and  explicit systematic methods for solving
special classes of difference equations are well known. For
instance the key to get the general solution of a linear
difference equation with constant coefficients, as for linear
differential equations once again with constant coefficients, is
to find generally the complex roots of an algebraic equation known
as the characteristic equation. To write down the general solution
at this point is only a trivial systematic matter. \par The
simplicity of the second-order difference equation
\begin{equation}Y_{p+2}=\mathcal{L}_0Y_p+\mathcal{L}_1Y_{p+1},\quad p=0,1,2,...=\mathbb{N}\end{equation}
   is, however, only
apparent when $\mathcal{L}_0,\mathcal{L}_1$, supposed independent
from $p$ and then behaving as constant coefficients, are
noncommutative \lq\lq mathematical objects\rq\rq. The specific
nature of these coefficients, as well as of $Y_p$ and of the same
\lq\lq operations\rq\rq appearing in equation (1) is of course
strictly related with the \lq\lq abstract support space\rq\rq
\hspace{0.1cm} in which the scientific problem represented by the
equation under scrutiny is formulated. Thus, for example, let our
reference space be an $n$-dimensional linear vector space $M$,
then $Y_p$ is a vector belonging to $M$, while
$\mathcal{L}_0,\mathcal{L}_1$ are linear operators acting upon the
vectors of $M$. Equation (1)  may even represent a matrix
equation, interpreting both the unknowns $Y_p$ and the
coefficients $\mathcal{L}_0,\mathcal{L}_1$ as matrices of given
order[1].
\par If $x\in \mathbb{R}$ and $ Y_p\equiv f_p(x)$ is a $C^\infty-$
function, then $\mathcal{L}_0,\mathcal{L}_1$ might be
noncommutative operators involving for example derivatives of any
order.
\par In  the context of quantum mechanics the unknowns of equation (1) might be
elements of a sequence of operators belonging to the Hilbert vector
space of the physical system  and $\mathcal{L}_0,\mathcal{L}_1$
 appropriate noncommutative superoperators [2,3]. \par These simple examples
 are sufficient to motivate the interest toward  the search  of techniques for
 solving equation (1) when $\mathcal{L}_0,\mathcal{L}_1$ are noncommutative
 coefficients. In this paper  we cope with  a Cauchy problem  associated
 with equation (1) and succeed in giving its
explicit solution  independently on the scientific context in
which equation (1) has been formulated. \\ We do not choose from
the very beginning the mathematical nature
 of its ingredients, rather we only require that all the symbols
 and operations appearing in equation (1) are meaningful. Thus
 \lq\lq
 vectors\rq\rq \hspace{0.1cm} $Y$ may be added, this operation being commutative and
 at the same time may be acted upon by $\mathcal{L}_0$ or
 $\mathcal{L}_1$ ( hereafter called operators) transforming themselves into other \lq\lq vectors\rq\rq. \\The
 symbol $Y_0=0$ simply denotes, as usual, the neutral elements of
 the addition between vectors. Finally we put $(\mathcal{L}_a\mathcal{L}_b)Y\equiv \mathcal{L}_a(\mathcal{L}_bY)\equiv
 \mathcal{L}_a\mathcal{L}_bY$ with $a$ or $b=0,1$ and define
 addition between operators through linearity.
\par The paper is organized as
follows. \\Some mathematical preliminaries open the next section
in which the Cauchy problem associated with equation (1) is
formulated. In the same section we state  and demonstrate our main
result which is the formula for solution of the Cauchy problem. In
the subsequent section we show the reduction of our formula to the
case of commutative coefficients by discussing a specific example.
Some concluding remarks are presented in the last section.

\vspace{0.5cm} \begin{center}\textmd{2. THE CAUCHY PROBLEM:
DEFINITION AND RESOLUTION}
\end{center}\vspace{0.2cm} \par We start by introducing some mathematical preliminaries.
\\ Let $u\in\mathbb{N}$ and $v\in \mathbb{N}$ and consider the
couple of generally noncommutative coefficients $\mathcal{L}_0$
and
 $\mathcal{L}_1$ appearing in equation (1).\\ We introduce the
symbol $\{\mathcal{L}_0 ^{(u)}\mathcal{L}_1^{(v)}\}$  to denote
the sum of all possible distinct permutations of $u$ factors
$\mathcal{L}_0$  and $v$ factors
 $\mathcal{L}_1$. A few examples help to illustrate  the notation:
\begin{center}$\{\mathcal{L}_0
^{(0)}\mathcal{L}_1^{(0)}\}=I,\{\mathcal{L}_0
^{(0)}\mathcal{L}_1^{(1)}\}=\mathcal{L}_1$\end{center}
\begin{center}$\{\mathcal{L}_0
^{(1)}\mathcal{L}_1^{(1)}\}=\mathcal{L}_0
 \mathcal{L}_1+\mathcal{L}_1\mathcal{L}_0$\end{center}\begin{center}$\{\mathcal{L}_0 ^{(1)}\mathcal{L}_1^{(2)}\} =\mathcal{L}_0 \mathcal{L}_1^2+\mathcal{L}_1 ^2\mathcal{L}_0+\mathcal{L}_1\mathcal{L}_0 \mathcal{L}_1$\end{center}
 When we exploit combinatorial theory, it is not difficult to convince ourselves that the number of all the possible different terms appearing in $\{\mathcal{L}_0 ^{(u)}\mathcal{L}_1^{(v)}\}$ coincides with the binomial coefficient $ \left(\begin{array}{c}
                                u+v\\ m
                                \end{array} \right)$, $m$ being the minimum between $u$ and
                                $v$ that is $m=min(u,v)$. \vspace{0.3cm}
\par We now consider the Cauchy problem
\begin{equation}\left\{\begin{array}{rl}
      Y_{p+2}=\mathcal{L}_0Y_p+\mathcal{L}_1Y_{p+1}\\
       Y_0=0,Y_1=\overline{Y}_1\end{array}\right., \quad p\in \mathbb{N},\end{equation}
       with  $\mathcal{L}_0$  and $\mathcal{L}_1$
        in general  noncommutative coefficients. \\Then we state the
        following
\begin{theorem}The solution of the Cauchy problem (2) can be
written as
\begin{equation} Y_p=\sum\limits_{t=0}^{\overline{t}_p}\{\mathcal{L}_0^{(t)}\mathcal{L}_1^{(p-1-2t)}\}\overline{Y}_1, \end{equation}
where \begin{equation} \overline{t}_p=\left[\frac{p-1}{2}
\right]=\left\{\begin{array}{rl}
      \frac{p-2}{2}, \quad if \quad p  \quad even\\
       \frac{p-1}{2}, \quad if \quad p \quad odd\end{array}\right.,\end{equation}
 denoting by $[x]$  the integer part of $x\in \mathbb{R}$.
\end{theorem}
\emph{Proof}:\\ In order to prove that equation (3) gives the
solution of the Cauchy problem (2) we exploit the  procedure of
mathematical induction.
\par To this end we put $p=1$ in equation (3). It is immediate to
verify that, in this case, equation (3) verifies the initial
condition $Y_1=\overline{Y}_1$. \\ We now consider the case $p=2$.
The sum appearing in equation (3) reduces, in this case, to a
single term corresponding to $t=0$. We have indeed
\\$Y_2=\mathcal{L}_1(\overline{Y}_1)$. Once again it is easy to
convince ourselves that $Y_2$ satisfies equation (2).
\par We now suppose that the first $(p+1)$ terms of the sequence
solution
 may be represented by equation (3). We have to prove that $Y_{p+2}$ also can be expressed by equation (3).\\ We observe that by the
 inductive
 hypothesis\\
$\mathcal{L}_0Y_p+\mathcal{L}_1Y_{p+1}=\sum\limits_{t=0}^{\overline{t}_p}\mathcal{L}_0\{\mathcal{L}_0^{(t)}\mathcal{L}_1^{(p-1-2t)}\}\overline{Y}_1+\sum\limits_{t=0}^{\overline{t}_{p+1}}\mathcal{L}_1\{\mathcal{L}_0^{(t)}\mathcal{L}_1^{(p-2t)}\}\overline{Y}_1
=\\$
\begin{equation}\sum\limits_{t=0}^{\overline{t}_p}\mathcal{L}_0\{\mathcal{L}_0^{(t)}\mathcal{L}_1^{(p-1-2t)}\}\overline{Y}_1+
\sum\limits_{t=1}^{\overline{t}_{p+1}}\mathcal{L}_1\{\mathcal{L}_0^{(t)}\mathcal{L}_1^{(p-2t)}\}\overline{Y}_1+\mathcal{L}_1^{(p+1)}\overline{Y}_1\end{equation}
\vspace{0.3cm}

It is not difficult to see that equation (4) implies
\begin{equation} \overline{t}_{p+1}=\left\{\begin{array}{rl}
      \overline{t}_p+1, \quad if \quad p  \quad even\\
       \overline{t}_p,\quad if \quad p \quad odd\end{array}\right..\end{equation} The right member of the equation (5) can be thus  rewritten as \vspace{0.3cm}
\\ $\mathcal{L}_1^{(p+1)}\overline{Y}_1+\mathcal{L}_0^{(1)}\mathcal{L}_1^{(p-1)}\overline{Y}_1+\mathcal{L}_1\{\mathcal{L}_0^{(1)}\mathcal{L}_1^{(p-2)}\}\overline{Y}_1+
\mathcal{L}_0\{\mathcal{L}_0^{(1)}\mathcal{L}_1^{(p-3)}\}\overline{Y}_1+\vspace{0.2cm}
\\\mathcal{L}_1\{\mathcal{L}_0^{(2)}\mathcal{L}_1^{(p-4)}\}\overline{Y}_1+\dots+
\mathcal{L}_0\{\mathcal{L}_0^{(t)}\mathcal{L}_1^{(p-1-2t)}\}\overline{Y}_1+\mathcal{L}_1\{\mathcal{L}_0^{(t+1)}\mathcal{L}_1^{(p-2(t+1)
)}\}\overline{Y}_1+$\vspace{0.2cm}\begin{equation}\dots
+\left\{\begin{array}{rl}
      \mathcal{L}_0\{\mathcal{L}_0^{(\overline{t}_p)}\mathcal{L}_1^{(p-1-2\overline{t}_p-1)}\}\overline{Y}_1+\mathcal{L}_1\{\mathcal{L}_0^{(\overline{t}_p +1)}\mathcal{L}_1^{(p-2(\overline{t}_p+1))
}\}\overline{Y}_1,\quad if \quad p  \quad even\\
      \mathcal{L}_0\{\mathcal{L}_0^ {(\overline{t}_p)}\mathcal{L}_1^{(p-1-2\overline{t}_p)}\}\overline{Y}_1=\mathcal{L}_0^{\frac{p+1}{2}}\overline{Y}_1,\quad if \quad p \quad
      odd\end{array}\right.\end{equation}

\par Concentrating now  on the general $t^{th}-$ operator term
of equation (7), that is
\begin{equation}
\mathcal{L}_0\{\mathcal{L}_0^{(t)}\mathcal{L}_1^{(p-1-2t)}\}+\mathcal{L}_1\{\mathcal{L}_0^{(t+1)}\mathcal{L}_1^{(p-2(t+1))
}\},\quad  t\in [0,\overline{t}_p)\cap\mathbb{N},\end{equation} it
is not difficult to persuade ourselves that
 by definition it expresses the sum of $\left(\begin{array}{c}
                                p-t-1\\ m
                                \end{array} \right)+\left(\begin{array}{c}
                                p-t-1\\ m'
                                \end{array} \right)$ terms,  where
                                $m=min(t,p-2t-1)$ and
                                $m'=min(t+1,p-2t-2)$.
  \vspace{0.2cm}  \\ It is possible to verify that, when  $t\in [0,\overline{t}_p)\cap\mathbb{N}$, then $m\geq0, m'\geq0$ and
     $ |m-m'|=1$ so that in according with the well-known Stifel formula
\begin {equation}\left(\begin{array}{c}
                                p-t-1\\ m
                                \end{array} \right)+\left(\begin{array}{c}
                                p-t-1\\ m'
                                \end{array} \right)=\left(\begin{array}{c}
                                p-t\\ M
                                \end{array} \right),\end{equation}
 where $M=max(m,m')=min(t+1,(p+2)-1-2(t+1))$.

 \vspace{0.2cm}In view of relation (9) we may thus say that the
 expression (8) coincides with the sum of all possible distinct
 permutations of $(t+1)$ factors $\mathcal{L}_0$ and
 $[(p+2)-1-2((t+1)]=p-1-2t$  factors $\mathcal{L}_1$. Thus by
 definition we may legitimately  write  that
\begin{equation}
\mathcal{L}_0\{\mathcal{L}_0^{(t)}\mathcal{L}_1^{(p-1-2t)}\}+\mathcal{L}_1\{\mathcal{L}_0^{(t+1)}\mathcal{L}_1^{(p-2(t+1))
}\}=\{\mathcal{L}_0^{(t+1)}\mathcal{L}_1^{(p+2)-1-2(t+1)}\},
(\forall) t\neq\overline{t}_p\end{equation} \par We also observe
that the first term of equation (7) can be cast
 in the following form

\begin{equation} \mathcal{L}_1^{p+1}={\mathcal{L}_0^0\mathcal{L}_1^{(p+2)-1-2\cdot 0}}\end{equation}
The definition of $\overline{t}_p$ given by equation (4) on the
other hand enables us to say that
 \begin{equation}
\overline{t}_{p+2}=\left[\frac{p+1}{2}\right]=\left\{\begin{array}{rl}
      \overline{t}_p+1=\frac{p}{2}, \quad if \quad p  \quad even\\
       \overline{t}_p+1=\frac{p+1}{2}, \quad if \quad p \quad odd\end{array}\right..\end{equation}
Thus the last term of expression (7) may be written
as\begin{equation} \left\{\begin{array}{rl}
      \{\mathcal{L}_0^{(\overline{t}_p
      +1)}\mathcal{L}_1^{((p+2)-1-2(\overline{t}_p+1))
}\}, \quad if \quad p  \quad even\\
      \mathcal{L}_0^{\frac{p+1}{2}}, \quad if \quad p \quad
      odd\end{array}\right.=
\{\mathcal{L}_0^{(\overline{t}_{p+2})}\mathcal{L}_1^{((p+2)-1-2\overline{t}_{p+2})}\}\end{equation}
Using relations (7),(10),(11) and (13) we may thus conclude
that\\
 \begin{equation}\mathcal{L}_0Y_p+\mathcal{L}_1Y_{p+1}=\sum\limits_{t=-1}^{\overline{t}_{p+2}}\{\mathcal{L}_0^{(t+1)}\mathcal{L}_1^{((p+2)-1-2(t+1))}\}\overline{Y}_1=\sum\limits_{t=0}^{\overline{t}_{p+2}}\{\mathcal{L}_0^{(t)}\mathcal{L}_1^{((p+2)-1-2t)}\}\overline{Y}_1\end{equation}
which coincides with $Y_{p+2}$ in accordance with the resolving
formula given by equation (3). \hspace{10cm}$\Box $\vspace{0.5cm}
\begin{center}\textmd{3. REDUCTION TO THE CASE OF COMMUTATIVE COEFFICIENTS}\end {center}
\vspace{0.2cm}
\par It is well-known [4] that, when the coefficients $\mathcal{L}_0$ and
$\mathcal{L}_1$ of equation (1) reduce to C-numbers as well as
$Y_p$, the solution of the relative equation (1) is traced back to
the solutions of its characteristic equation. The following
theorem summarizes the well known result.
\begin{theorem} Consider the equation
\begin{equation}y_{p+2}=c_0y_p+c_1y_{p+1},\end{equation}
where $c_0\neq0$ and $c_1$ are real constants associated with  the
initial conditions $y_0=1$ and $y_1=\overline{y}_1$.

If $m_1$ and $m_2$ are the  roots of the characteristic equation
\begin{equation}m^2-c_1m-c_0=0, \quad m_{1,2}=\frac{c_1\pm
\sqrt{\Delta}}{2},\quad \Delta=c_1^2+4c_0,\end{equation} then the
solution of this Cauchy problem is given by
\begin{equation}
y_p= \left\{\begin{array}{rl}
      \frac{1}{m_1-m_2}[m_1^p-m_2^p]\overline{y}_1,\quad if \quad \Delta \neq 0\\
       pm_1^{p-1}\overline{y}_1,\quad if \quad \Delta=0\end{array}\right..\end{equation}
        .
\end{theorem}\vspace{0.2cm}

We now demonstrate that our operator solution (3) reduce to (17)
when $\mathcal{L}_0=c_0$ and $\mathcal{L}_1=c_1$. We firstly prove
the following

\begin{lema} When $\mathcal{L}_0=c_0$ and
$\mathcal{L}_1=c_1$, then $y_p$ given by equation (3) can be
reduced to the following expression
\begin{equation} y_p=\sum\limits_{t=0}^{\overline{t}_p}\{\mathcal{L}_0^{(t)}\mathcal{L}_1^{(p-1-2t)}\}\overline{y}_1=c_1^{p-1}\sum\limits_{t=0}^{\left[\frac{p-1}{2}\right]} \left(\begin{array}{c}
                                p-t-1\\ t
                                \end{array} \right) \left(\frac{c_0}{c_1^2}\right)^t \overline{y}_1\end{equation}

\end{lema}
\emph{Proof}: \par The solution of a linear homogenous
second-order difference equation with constant coefficients
presented in the form (18) is a consequence of the general result
presented in section 2. In this case the coefficients are
constants and so they commute. We may conclude that

\begin{equation} y_p=\sum\limits_{t=0}^{\overline{t}_p} \left(\begin{array}{c}
                                p-t-1\\ min\{t,p-1-2t\}
                                \end{array} \right)
                                c_0^tc_1^{p-1-2t}\overline{y}_1\end{equation}
Taking in consideration that
\begin{equation}\left(\begin{array}{c}
                                p-t-1\\ t
                                \end{array} \right) =\left(\begin{array}{c}
                                p-t-1\\ p-1-2t
                                \end{array} \right) \end{equation}
we may rewrite the equation (19) in the simplest way as (18).
$\Box$
                                \vspace{0.5cm}
\par  In order to cast the expression (18) into the form presented
in theorem 2 we  consider two cases, accordingly with the nullity
or non nullity of the discriminant of the characteristic equation
(16).
\par Assuming  $\Delta\neq 0$  we may use the identity[5,6]
\begin{equation}\sum\limits_{k=0}^{[n/2]}\left(\begin{array}{c}
                                n-k\\ k
                                \end{array} \right)z^k=2^{-n-1}(1+4z)^{-1/2}[(1+\sqrt{1+4z})^{n+1}-(1-\sqrt{1+4z})^{n+1}],\end{equation}
where
$z\in \mathbb{C},z=c_0/c_1^2$, to rewrite the equation (18)
as
\\$  y_p=c_1^{p-1}2^{-p}\left(1+\frac{4c_0}{c_1^2}\right)^{-1/2}\left[\left(1+\sqrt{1+\frac{4c_0}{c_1^2}}\right)^p-\left(1-\sqrt{1+\frac{4c_0}{c_1^2}}\right)^p\right]\overline{y}_1=$\\
$ \left(\frac{c_1}{2}\right)^{p}
\frac{1}{\Delta}\left[\left(\frac{c_1+\sqrt{\Delta}}{c_1}\right)^p-\left(\frac{c_1-\sqrt{\Delta}}{c_1}\right)^p\right]\overline{y}_1=$$\frac{1}{\sqrt{\Delta}}\left[\left(\frac{c_1+\sqrt{\Delta}}{2}\right)^p-\left(\frac{c_1-\sqrt{\Delta}}{2}\right)^p\right]\overline{y}_1=$
\par \begin{equation}=\frac{1}{m_1-m_2}\left[m_1^p-m_2^p\right]\overline{y}_1\end{equation} \vspace{0.3cm}

 When  on the other hand $\Delta =0$,  then $c_0/c_1^2=-1/4$ so
that exploiting  the identity[5,6]
\begin{equation}\sum\limits_{k=0}^{[n/2]}  \left(-\frac{1}{4}\right)^k\left(\begin{array}{c}
                                n-k\\ k
                                \end{array} \right)=(n+1)2^{-n}\end{equation}

we may easily get the following form for equation (18)

\begin{equation}y_p=\sum\limits_{t=0}^{[\overline{t}_p]}\left(-\frac{1}{4}\right)^t\left(\begin{array}{c}
                                p-1-t\\ t
                                \end{array}
                                \right)\overline{y}_1=p\left(\frac{c_1}{2}\right)^{p-1}\overline{y}_1=pm_1^{p-1}\overline{y}_1.\end{equation}

\vspace{0.5cm}
\begin{center}\textmd{CONCLUDING REMARKS}\end {center}\vspace{0.2cm}
The main result of this paper is that expressed by equation (3).
It provides the exact solution of the Cauchy problem formulated in
equation (2). At the best of our knowledge this resolutive
formula, the peculiar feature of which is its wide applicability,
is a new result demonstrated in this paper for the first time. The
importance of being able to treat in a systematic way mathematical
problems, wherein effects stemming from noncommutativity cannot be
simply overcome, is very easy to appreciate and in addition is not
confined to few research areas. For all these reasons we claim
that our results are both interesting and useful.  \vspace{0.1cm}
\begin{center}\textmd{REFERENCES}\end{center}
1. Gantmacher F.R. \textit{The Theory of Matrices} American
Mathematical \par Society,  Providence, Rhode Island, 1998.
\\2. Breuer H.-P., Petruccione F. \textit{The Theory of Open
Quantum Systems}\par Oxford University Press Inc., New York, 2002.
\\3. Le Bellac M. \textit{Quantum Physics} Cambridge University
Press, Cambridge,\par 2006.
\\4. Kelley W. G.  Peterson A.C. \textit{ Difference Equation}, Academic Press,\par London, 2001.\\
5. Prudnikov A.P. Brychov Yu. A. Marichev O.I.\textit{ Integrals
and Series }\par vol 1, Gordon and Breach Science Publishers,
Amsterdam, 1990.
\\ 6. Graham R. Knuth D. Patashnik O.
\textit{Concrete Mathematics}, Addison-\par Wesley Publishing
Company, Reading, Massachusetts, 1994.

\end{document}